\documentclass[reprint,prb,aps,groupedaddress,amsmath,amsfonts]{revtex4-1}
\usepackage{mathtools}
\usepackage{graphicx}

\DeclareMathOperator{\sech}{sech}
\DeclareMathOperator{\csch}{csch}

\begin{document}

\title{Bounds on quantum confinement effects in metal nanoparticles}

\author{G. Neal \surname{Blackman} III}
\affiliation{College of Engineering and Science, Louisiana Tech University, Ruston, Louisiana 71272, USA}

\author{Dentcho A. \surname{Genov}}
\affiliation{College of Engineering and Science, Louisiana Tech University, Ruston, Louisiana 71272, USA}
\email{dgenov@latech.edu}
\date{\today}

\begin{abstract}
Quantum size effects on the permittivity of metal nanoparticles are investigated using the quantum box model. Explicit upper and lower bounds are derived for the permittivity and relaxation rates due to quantum confinement effects. These bounds are verified numerically, and the size-dependence and frequency-dependence of the empirical Drude size parameter is extracted from the model. Results suggest that the common practice of empirically modifying the dielectric function can lead to inaccurate predictions for highly uniform distributions of finite-sized particles.
\end{abstract}


\maketitle

\section{Introduction}

Metallic nanostructures and their associated surface plasmon resonances (SPRs) enable incident light to be intensified by several orders of magnitude. This plasmonic enhancement plays a paramount role in a broad range of emerging optical technologies, including electromagnetic cloaks and metamaterials, \cite{Pendry:2006hq, Leonhardt:2006ex, Cai:2007jg, Shalaev:2007tz} superlenses, \cite{Pendry:2000wg, Cai:2005kg, Nielsen:2010vc} ultrafast optoelectronics,\cite{Ozbay:2006dz, Zia:2006ia, MacDonald:2008fi} cancer treatments,\cite{Huang:2007gj,Conde:2012ir} and sensitive chemical sensors.\cite{Kneipp:1999kg,Genov:2004iv,Li:2015ed} The mechanism of plasmonic enhancement has been well studied within the context of classical electrodynamics. However, as the feature size of nanomaterials approach atomic scale, quantum-mechanical effects emerge, which in some cases can further enhance their plasmonic properties. In this paper, we demonstrate that bounds can be placed on the additional effects due to quantum confinement.

The optical behavior of large metal nanoscale objects (dimensions $\gtrsim$ 10 nm) are described using bulk permittivity functions and classical electrodynamics methods such as Mie theory \cite{Mie:1908tn} or the discrete-dipole approximation.\cite{Flatau:1994cz, Schatz:1995tl, Kelly:2003jj} Bulk permittivity functions of metals $\varepsilon(\omega) = \varepsilon_b(\omega) + \varepsilon_D(\omega)$ can be formally separated into the bound electron contribution $\varepsilon_b(\omega)$ and the conduction electron (Drude) contribution $\varepsilon_D(\omega)$ with the Drude component given by \cite{Mermin:1976ud, Drachev:2008icba}
        \begin{align}
        \label{eq:drude_fn}
            \varepsilon_D(\omega) &= \varepsilon_D'(\omega) + i \varepsilon_D''(\omega) \nonumber \\ &= 1 - \frac{\omega_{p}^2}{\omega^2+ \gamma^2} + i\frac{\omega_{p}^2 \gamma}{\omega(\omega^2 + \gamma^2)},
        \end{align}
where $\omega_{p} = ({n_e e^2} / {\varepsilon_0 m_e})^{1/2}$ is the plasma frequency, which depends on the conduction electron density $n_e$, the electron charge $e$, and the effective electron mass $m_e$; $\gamma$ is a phenomenological damping constant (relaxation rate) which equals the SPR bandwidth $\Gamma$ for a free electron gas in the limit $\omega \gg \gamma$.\cite{Vollmer:1995vu} For the majority of this paper, we will focus on the conduction band contribution since it is dominant in the infrared and visible frequencies for many metals. We will revisit the bound electron component in the section dedicated to experimental comparison since it plays an important role in noble metals like silver, copper, and gold at frequencies close to the surface plasmon resonance frequency.\cite{Christy:1972di}

Although Eq.~(\ref{eq:drude_fn}) accurately describes the conduction electron behavior in bulk metals and large metal particles, the bulk Drude response is inadequate for tiny metal particles with radii less than 10 nm. At this size range, the particle surface starts to play an important role in the optical response, and the optical functions become size dependent.\cite{Kreibig:1969ja,Vollmer:1995vu} In contrast to Drude theory, which predicts constant damping rates, optical measurements have established that damping rates are inversely proportional to the particle radius when $R < 10$~nm.\cite{Doremus:1965ih,Hovel:1993jo,Vollmer:1995vu} Thus there has been great success in reproducing experimental measurements by replacing the relaxation rate in Eq.~(\ref{eq:drude_fn}) with the size dependent term\cite{Kreibig:1969ja,Genzel:1985ub,Vollmer:1995vu,El-Sayed:1999dr,Baida:2009hl}
        \begin{equation}
            \gamma(R) = \gamma_0 + \mathcal{A} \frac{v_{F}}{R}
        \label{eq:sizeparam}
        \end{equation}
where $v_F$ is the Fermi velocity, $\gamma_0$ is the bulk value of the damping constant, and $\mathcal{A}$ is an empirical size parameter of the order of 1.

The $1/R$ contribution in Eq. (\ref{eq:sizeparam}) has been derived within the context of several different theoretical frameworks, each with their own interpretation of the size-dependent damping rates. The classical \textit{free path effect}\cite{Kreibig:1969ja} considers it a result of increased surface scattering in finite-size particles, which leads to a modified mean free path of the electrons. Semiclassical and nonlocal models interpret the modified damping rates to be a consequence of electron-hole pair formation (Landau damping) and surface screening in hydrodynamic models.\cite{Broglia:1992vi,Raza:2011io,Raza:2015ef,Penn:1983ig,Monreal:2013ea,Monreal:2014vx} The canonical particle-in-a-box model pioneered by Kawabata and Kubo\cite{Kawabata:1966bt} and improved subsequently by numerous others\cite{Hache:1986tx, Rice:1973it, Lushnikov:1973ca, Genzel:1975wh, Yatom:1976em, Cini:1981to, Ashcroft:1982fq, Schatz:1983di, Barma:1989tx, Ivanyuk:2008bv, Lue:1994ej, Rautian:1997cf,  Govyadinov:2011ev} has shown that the size-dependent damping rates can also be considered a quantum size effect.

Although many theoretical approaches have derived Eq.~(\ref{eq:sizeparam}), the value of the size parameter has been debated in the literature, with values for $\mathcal{A}$ ranging from 0.1 to 2 depending on the details of the calculation performed.\cite{Genzel:1985ub,Vollmer:1995vu,Alvarez:1997kw} Further complicating matters, experiments have reported an even wider range of values,\cite{Quinten:1996uf,Hilger:2001it,Hilger:2005wa} showing that the proportionality constant is sensitive to a multitude of microscopic effects such as particle geometry, surface roughness, and chemical interface damping, all of which become significant when the particle radius is less than 10 nm.\cite{Halperin:1986jd, Vollmer:1995vu}

In principle, many detailed effects could be considered by first-principle investigations. However, theoretical descriptions of metal particles based on \textit{ab initio} techniques and time-dependent (TD) density functional theory (DFT)\cite{Bonacic-Koutecky:1991uv, Parrinello:1991uj} are usually restricted to particles consisting of less than 120 atoms because of their computational demand.\cite{Idrobo:2007bs, Bonacic-Koutecky:2001bk, Baishya:2008iy, Aikens:2008dw} More optimized techniques include less physical details, but even these methods are limited to particles smaller than  $R \approx$ 2.5 nm.\cite{Kuisma:2015vc, He:2010hi, Jensen:2009hw} Because of practical limitations, these theories have limited usefulness for modeling general plasmonic systems; to model larger particles, simpler quantum-mechanical methods are required. These include the jellium model\cite{Brack:1993cs,deHeer:1993ue,Yannouleas:1993vy,Molina:2003ks,Weick:2005fz,Lermé:2010bi} and quantum box models (QBMs).

The primary advantage of the particle-in-a-box technique, referred to as the QBM, is that it uniquely allows for analytical solutions and closed-form expressions, which provide more physical insight than purely numerical schemes. We use the framework of the QBM to evaluate the limits to which quantum confinement of the conduction electrons can further enhance the optical properties of metal nanoparticles. The results also help establish the limitations of using Eq. (2) to empirically modify the Drude function. We focus primarily on damping effects since line-shape broadening is the most pronounced size-dependent effect for SPRs of metal particles in the $R=1$--10 nm range.

In Sec. \ref{sec:qbm}, we introduce the spherical QBM and summarize numerical calculations for finite-sized systems. The results are consistent with previous finite-size calculations,\cite{Barma:1989tx,Molina:2002wf,Molina:2003ks,Govyadinov:2011ev} showing that the permittivity $\varepsilon$ and the relaxation rate $\gamma$ are characterized by a range of values that fluctuate sensitively with respect to frequency and particle size. We demonstrate that the Drude size parameter $\mathcal{A}$ is similarly characterized by fluctuating values since it is directly linked to $\gamma$. Hence, in Sec. \ref{sec:bounds}, we derive analytical bounds on the fluctuations in these three related quantities, and we verify the analytical bounds using numerical calculations. Finally, in Sec. \ref{sec:comparison_with_experiment}, we compare experimental measurements of the size parameter with numerical QBM calculations for nonuniform size distributions. Results point toward a different approach for treating the optical response of nano-sized metallic systems. Specifically, we argue in Sec. \ref{sec:conclusion} that for highly uniform size distributions, either: (i) a quantum-mechanical permittivity calculation should be used, (ii) a frequency- and size-dependent function $\mathcal{A}(\omega, R)$ must be introduced, or (iii) the proper bounds should be used to calculate the range of expected optical properties.

\section{Quantum box model (QBM)\label{sec:qbm}}
\begin{figure}
\includegraphics{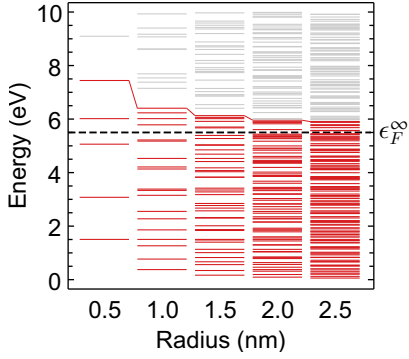}
\caption{\label{fig1}(Color online) Energy levels of an infinite spherical well with varying radius, representing silver nanospheres with a constant electron density $n_e=5.86 \times 10^{28}$ $\text{m}^{-3}$. The occupancy of each energy level is filled according to its degeneracy factor $2(2l+1)$. Dark (red) lines indicate occupied states at temperature $T=0$ K, and light (gray) lines are unoccupied.}
\end{figure}

In the QBM, $N$ conduction electrons are confined within an infinite potential well whose dimensions are designed to represent a particle of the same size. The electrons are assumed to be non-interacting, with each electron belonging to its own single-electron eigenstate. In a spherical well of radius $R$, the wavefunctions have the form $\psi_{nlm} = A_{nl} j_l(a_{nl}r/R)Y_{l}^{m}(\theta, \phi)$ where $A_{nl} = (2/R^3)^{1/2} / j_{l+1}(a_{nl})$ is the normalization constant, $Y_{l}^{m}$ denotes the spherical harmonics ($-l \leq m \leq l$), and $a_{nl}$ represents the $n$th zero of the spherical Bessel function $j_l$ with order $l \geq 0$. These states have the energy levels $E_{nl} = \hbar^2a_{nl}^2/2m_eR^2$ with degeneracy $g(E_{nl})=2l+1$ and are presented in Fig.~\ref{fig1}. The electromagnetic response of the conduction electrons is given by the standard quantum mechanical susceptibility tensor for linear materials:\cite{Boyd:2008tl}
\begin{equation}
	\label{eq:qm_susceptibility}
    \chi^{ij}(\omega, R) = \frac{1}{\varepsilon_0 \hbar V} \sum_s{\sum_{s'}{ \frac{(\textsl{w}_s-\textsl{w}_{s'})\mu_{ss'}^i\mu_{s's}^j}{\omega_{ss'}-\omega-i\gamma_{ss'}/2} } }\text{,}
\end{equation}
where $\textsl{w}_s$ and $\textsl{w}_{s'}$ are the occupation numbers of the states $\left| s \right\rangle$ and $\left| s' \right\rangle$, $\omega_{ss'} = (E_{s'} - E_s)/\hbar$ is the transition frequency, $V$ is the volume of the particle, and $\gamma_{ss'}$ is the transition relaxation rate. The electric dipole transition moments are given by $\mu_{ss'}^j = \left\langle s \left| e r_j \right| s' \right\rangle$ where $r_j$ is the displacement in the direction of the polarization unit vector $\hat{\text{e}}_j$. Note that Eq.~(\ref{eq:qm_susceptibility}) does not account for additional effects due to inhomogeneity of the electron density near the particle surface (non-locality), which is expected to play a role in very small particles with dimensions comparable to $1/k_F$ where $k_F$ is the Fermi wavevector.\cite{Ashcroft:1982fq} These effects are best described by more detailed approaches like TD-DFT. For this reason, we restrict our analysis to particles with $R > 1$~nm.

The temperature dependence of the occupation numbers is given by Fermi statistics
\begin{equation}
\label{eq:fermi_dirac}
    \textsl{w}_s = \frac{2}{e^{(E_s - \epsilon_F)/k_B T  } + 1},
\end{equation}
where $k_B$ is Boltzmann's constant, $\epsilon_F$ is the Fermi energy, and the factor of two accounts for spin degeneracy of the electron. The Fermi energy can be considered a constant value for bulk material ($\epsilon_F^{\infty}$), but it should be treated as a size-dependent parameter for small particles when quantization of energy levels becomes important. We calculate the size-dependent Fermi energy $\epsilon_F(R)$ using an electron-counting process as follows. The occupancy of each energy state is determined by its degeneracy factor $2(2l+1)$, which includes both angular and spin-degeneracy. The energy levels are then filled from the ground state upward until we have accounted for each of the nanoparticle's $N=n_eV$ conduction electrons. At this point, the size-dependent Fermi energy has been reached. A visualization of this process can be seen in Fig.~\ref{fig1} where $\epsilon_F(R)$ approaches the bulk value as the particle size increases.

Wood and Ashcroft\cite{Ashcroft:1982fq} demonstrated that high-symmetry systems have an effective energy gap that scales with the number of electrons according to $\delta_E \sim \epsilon_F / N^{1/3}$. According to the Wood and Ashcroft scaling, typical energy gaps in the size range we consider ($R \leq 10$ nm) are $\delta_E \gtrsim 0.09$ eV. This was also confirmed numerically by looking at the distribution of dipole transition frequencies for particles in the R = 1--10 nm size range (see Fig. 1). Thus, within this size range, temperature effects play a minimal role since thermal energy fluctuations at ambient temperatures are of the order of $k_B T \approx 0.02$ eV. Therefore, we take $T=0$ K as a good approximation to the low temperature limit ($\delta_E \gg k_B T$). In this case, all states below the Fermi energy ($E \leq \epsilon_F$) are occupied with $\textsl{w}_s = 2$, and all states with $E > \epsilon_F$ are unoccupied with $\textsl{w}_s = 0$. Therefore, at zero temperature, the permittivity $\varepsilon(\omega) = 1 + \chi(\omega)$ becomes a sum over occupied ($o$) and unoccupied ($u$) states
\begin{equation}
\label{eq:susceptibility}
    \varepsilon^{ij}(\omega, R) = 1 + \omega_{p}^2 \sum_s^o{\sum_{s'}^u{ \frac{S^{ij}_{ss'}}{\omega_{ss'}^2-\omega^2-i\omega\gamma_{ss'}} } },
\end{equation}
where we have introduced the oscillator strengths $S^{ij}_{ss'}=4 m_e\omega_{ss'} \mu_{ss'}^i\mu_{s's}^j / \hbar N$, which satisfy the sum rule $\sum_{ss'}{S^{ii}_{ss'}}~=~1$. As verification of the low-temperature approximation, calculations have also been performed at $300 $ K (not shown here), which reveal negligible differences to the calculations performed at absolute zero.

\begin{figure*}
\includegraphics{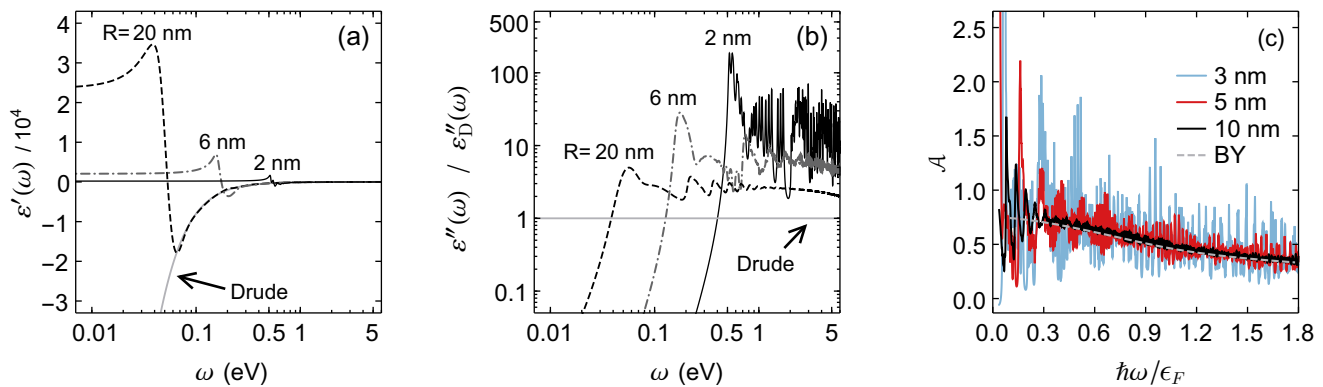}
\caption{\label{fig:permittivity_size_range} The (a) real part and (b) imaginary part of the permittivity of silver nanospheres calculated using the QBM. Convergence to the bulk Drude function can be seen as particle size increases. (c) The frequency dependence of $\mathcal{A}(\omega, R)$ extracted from the QBM permittivity using Eq. (\ref{eq:effective_size_param}). For comparison, the analytical result from Ref. \onlinecite{Broglia:1992vi} (BY) is shown in (c). Values used for silver were $\omega_{p}=9.1$ eV, $\gamma_0 = 0.021$ eV, and $n_e=5.86 \times 10^{28}$ $\text{m}^{-3}$ (constants obtained from Ref. \onlinecite{Christy:1972di} data).}
\end{figure*}

Because all directions are equivalent for spherical particles, we can choose the $z$-direction to coincide with the direction of polarization. The oscillator strengths $S^{zz}_{ss'}\equiv 4 m_e\omega_{ss'} \left| \left\langle s \left| z \right| s' \right\rangle \right|^2 / \hbar N$ are then evaluated between the initial state $\left| s \right\rangle = \left| \psi_{nlm} \right\rangle$ and final state $\left| s' \right\rangle  = \left| \psi_{n'l'm'} \right\rangle$. This leads to the selection rules $\Delta m = m' - m = 0$ and $\Delta l = l' - l = \pm 1$. Finally, the oscillator strengths for the allowed transitions are given by
\begin{equation}
    S^{zz}_{ss'} = \delta_{\Delta m,0} (C_{l+1}^m \delta_{\Delta l,1} + C_l^m \delta_{\Delta l,-1} ) \frac{16(a_{nl} a_{n'l'})^2}{N (a_{nl}^2 - a_{n'l'}^2)^3}
\end{equation}
where $C_l^m = (l^2 - m^2)/(4l^2-1)$. Since the energy levels do not depend on the quantum number $m$, it essentially represents a degeneracy factor that can be incorporated into the strength factors. Thus it is convenient to define new oscillator strengths $S^{zz}_{n,l,n',l'}=\sum_{m=-l}^l{S^{zz}_{ss'}}$ that are independent of $m$. Evaluating the sum over $m$, we find
\begin{equation}
\label{eq:exact_sif_2}
    S^{zz}_{n,l,n',l'} = ( \delta_{\Delta l,1} + \delta_{\Delta l,-1} ) \frac{16 (a_{nl} a_{n'l'})^2(l + l' + 1)}{3 N (a_{n'l'}^2 - a_{nl}^2)^3}\text{.}
\end{equation}
Equations (\ref{eq:susceptibility}) and (\ref{eq:exact_sif_2}) can then be used to write the permittivity as a sum over the quantum numbers $n$ and $l$:
\begin{align}
\label{eq:exactsum}
    \varepsilon(\omega, R) = 1 + & \frac{16 \omega_{p}^2}{3N}
    \sum_{nl}^o{ \sum_{n'l'}^u{
        \frac{(a_{nl} a_{n'l'})^2} {(a_{n'l'}^2 - a_{nl}^2)^3} }} \nonumber \\
        & \times \frac{ ( \delta_{\Delta l,1} + \delta_{\Delta l,-1} ) (l + l' + 1) }{\omega_R^2(a_{n'l'}^2-a_{nl}^2)^2-\omega^2-i\omega\gamma_{0}} 
    ,
\end{align}
where we have defined the size-dependent frequency $\omega_R = \hbar / 2 m_e R^2$. In Eq. (\ref{eq:exactsum}), we have suppressed the superscript $z$ since the direction is unimportant, and we have fixed the transition relaxation rates to that of the bulk $\gamma_{ss'}=\gamma_0$. In reality, $\gamma_{ss'}$ represents the natural decay rate of the transitions, but no direct measurement has been made of this quantity, and so we have followed the common practice of relating it to the conductivity relaxation rate.\cite{Schatz:1983di,Drachev:2004ey,Vollmer:1995vu}

The sum in Eq.~(\ref{eq:exactsum}) is then evaluated numerically over all possible transitions from occupied states to unoccupied states until a reasonably high accuracy is achieved.\cite{Govyadinov:2011ev} To monitor convergence, we enforced the sum rule with a very small tolerance $1 - \sum{S_{ss'}} < 10^{-4}$, which was achieved by including as many as 1.65 million transitions for the largest particle size considered in this study ($R = 20$ nm). Fewer transitions are required to achieve the same convergence for smaller particles.

Figures \ref{fig:permittivity_size_range}(a) and \ref{fig:permittivity_size_range}(b) show the results for several sizes of silver nanoparticles. Both the real and imaginary parts approach bulk Drude behavior for large particles, but discrete resonances are prominent for smaller sizes. Quantum effects are easily seen in the infrared region of $\varepsilon'$ where the QBM predicts that metal (silver) colloids/composites containing particles with radii less than 2 nm should have dielectric behavior ($\varepsilon' > 0$) for electromagnetic radiation with wavelengths larger than 2 microns (0.6 eV). Size effects are also noticeable in the visible frequency range where a decrease in particle size leads to a rapid increase of $\varepsilon''$ and the appearance of strong resonances.

We now consider the relaxation phenomena as predicted by the QBM. Although Eq.~(\ref{eq:sizeparam}) is sometimes studied only at the surface plasmon frequency, we consider the more general case by defining $\mathcal{A}$ to be the finite-size correction to the Drude permittivity at all frequencies. In the optical frequency range, we can assume $\omega \gg \gamma$, in which case we can define an effective relaxation frequency in terms of the imaginary part of the permittivity $\gamma(\omega, R) = \omega^3 \varepsilon''(\omega, R)/ \omega_p^2$. The effective size parameter then follows from Eq. (2) and is given by
\begin{equation}
\label{eq:effective_size_param}
\mathcal{A}(\omega, R) = \frac{\gamma(\omega, R) - \gamma_0}{v_F / R}.
\end{equation}
The frequency dependence of $\mathcal{A}(\omega, R)$ for fixed particle sizes is presented in Fig. \ref{fig:permittivity_size_range}(c), showing that the QBM predicts values for the size parameter that fluctuate above and below the smooth asymptotic result due to Barma and Subrahmanyam (later improved by Yannouleas and Broglia).\cite{Barma:1989tx, Broglia:1992vi} The common practice of extending the Drude model to be size-dependent [i.e. $\varepsilon_D(\omega, R)$] by combining Eqs.~(\ref{eq:drude_fn}) and (\ref{eq:sizeparam}) is only reasonable for high frequencies and large particle sizes where these deviations subside or when significant inhomogeneous broadening effects are present.

Considering that experimentally generated metal nanoparticle colloids and composite materials often consist of particles with sizes in the range of 1--10 nm, it is clear that quantum confinement effects are expected to play an important role for infrared and visible frequencies. In stark contrast to the semiclassical model, Fig. \ref{fig:permittivity_size_range} demonstrates that a small change in particle size or frequency can drastically change the value of the optical functions and size parameter. This exemplifies the importance of gaining a better understanding of finite-size effects in nanoscopic systems. Accordingly, the following section is dedicated to deriving bounds on the fluctuations predicted by the QBM.

\section{\label{sec:bounds}Analytical bounds on the permittivity and relaxation rate}

Previous authors have evaluated Eq.~(\ref{eq:exactsum}) or its equivalents by replacing summations with integrals,\cite{Kawabata:1966bt, Hache:1986tx, Barma:1989tx} which in effect smooths the resonances and averages the values of Fig. \ref{fig:permittivity_size_range}. These smoothing techniques approximate highly disperse experimental samples, but they conceal the full potential of what might be detected experimentally. Hence we take a different approach by seeking explicit bounds on the resonance behavior so that a range of expected values can be estimated when calculating optical properties.
	
We begin by obtaining broad bounds on the particle permittivity given by Eq.~(\ref{eq:susceptibility}) by minimizing and maximizing the summand with respect to $\omega_{ss'}$. The real part $\varepsilon'(\omega)$ has a minimum at $\omega_{ss'} = \sqrt{\omega(\omega - \gamma_0)}$ and maximum at $\omega_{ss'} = \sqrt{\omega(\omega + \gamma_0)}$. The imaginary part $\varepsilon''(\omega)$ has a maximum at $\omega_{ss'} = \omega$. Evaluating Eq.~(\ref{eq:susceptibility}) under these conditions and applying the oscillator strength sum rule, we readily obtain the following bounds for the permittivity
\begin{align}
	\label{eq:general_bounds}
	-\frac{ \omega_{p}^2 } {2 \omega \gamma_0} \leq \varepsilon'(\omega) - 1 \leq \frac{ \omega_{p}^2 } {2 \omega \gamma_0}, \quad
	0 \leq \varepsilon''(\omega) \leq \frac{ \omega_{p}^2 } {\omega \gamma_0} \text{.}
\end{align}
Although these bounds hold for all frequencies and particle sizes, they are not tight for high frequencies since they scale as $1/\omega$, which does not match the behavior of the Drude model for large $\omega$ ($\varepsilon_D' \sim 1/\omega^2$ and $\varepsilon_D'' \sim 1/\omega^3$). Furthermore, it is of practical value to find size-dependent bounds that capture quantum size effects. In the remainder of this section we seek size-dependent bounds with Drude-like behavior so that we can better characterize the effects of quantum confinement.

Following a procedure similar to Kraus and Schatz,\cite{Schatz:1983di} we approximate the spherical Bessel zeros using $a_{nl} \approx \pi(n + 1 + l/2)$, which can be recognized as the leading term of McMahon's asymptotic formula\cite{Stegun:1972uh} modified for spherical Bessel zeros with $n\geq0$. This is the simplest method that, as shown below, allows for obtaining tighter analytical bounds. McMahon's formula is exact for $l = 0$, so we define $n_{F}$ as the value of the quantum number $n$ on the Fermi surface when $l=0$. With this definition, the Fermi energy is $\epsilon_{F}=\hbar\omega_R\pi^2(n_{F}+1)^2$, and the Fermi surface is defined by the line $l=2(n_{F}-n)$. Occupied states lie below this surface where $0 \leq n \leq n_F$, $0 \leq l \leq 2(n_F - n)$, and $-l \leq m \leq l$. Summing the occupied states and keeping only the leading term for large $n_F$, we find the relation between $n_F$ and the number of states, $N_s = (4/3) n_F^3$. Taking the states to be doubly occupied ($N_s = N/2$) and inserting the approximate $a_{nl}$, the oscillator strengths follow from Eq.~(\ref{eq:exact_sif_2})
\begin{align}
	\label{eq:approx_sif}
    S_{ss'} = \thinspace &S_{n,l,\Delta n, \Delta l} \equiv (\delta_{\Delta l,1} + \delta_{\Delta l,-1})
        \frac{8(2n+l+2)^2}{9 n_F^3 \pi^2(2\Delta n + \Delta l)^3} \nonumber \\
        & \times \frac{(2l+\Delta l+1)(2n+\Delta n+l+\Delta l + 2)^2}{(4n + 2\Delta n + 2l + \Delta l + 4)^3} \text{.}
\end{align}
Using the approximate energies $E_{nl} = \hbar^2 \pi^2 (n + 1 + l/2)^2 / 2 m_e R^2$, the transition frequencies are
\begin{equation}
	\label{eq:approx_wif}
	\omega_{n,l,\Delta n, \Delta l} \equiv \frac{\pi^2 \omega_R}{4} (2\Delta n + \Delta l)(2\Delta n + \Delta l + 4 + 2l + 4n)
\end{equation}
Therefore, the permittivity under the McMahon approximation is
\begin{equation}
\label{eq:eps_mcmahon}
    \varepsilon(\omega, R) = 1 + \omega_{p}^2 \sum_{
    	\mathclap{\substack{n,l \\ \Delta n, \Delta l}}
    }{ \frac{S_{n,l,\Delta n, \Delta l}}{\omega_{n,l,\Delta n, \Delta l}^2-\omega^2-i\omega\gamma_0} } \text{.}
\end{equation}
In writing Eqs. (\ref{eq:approx_sif})-(\ref{eq:eps_mcmahon}), we have used the transition notations $\Delta l = l' - l$ and $\Delta n = n' - n$. With this convention, the sum over states in Eq. (\ref{eq:susceptibility}) has become a sum over values of $\Delta n$ and $\Delta l = \pm 1$ for which the occupied states transition to an unoccupied state. This leads to the additional summation constraints
\begin{subequations}
\label{eq:sum_limits}
\begin{align}
	& \Delta n \geq 1 - \frac{1-\Delta l}{2}  \\
    & 0 \leq n \leq n_{F} - \frac{1-\Delta l}{2} \\
    & l \geq \text{max}\left[0,2\left(n_{F} - n - \Delta n + \frac{1-\Delta l}{2}\right)\right] \\
    & l \leq 2(n_{F} - n)
\end{align}
\end{subequations}
\begin{figure*}
\includegraphics{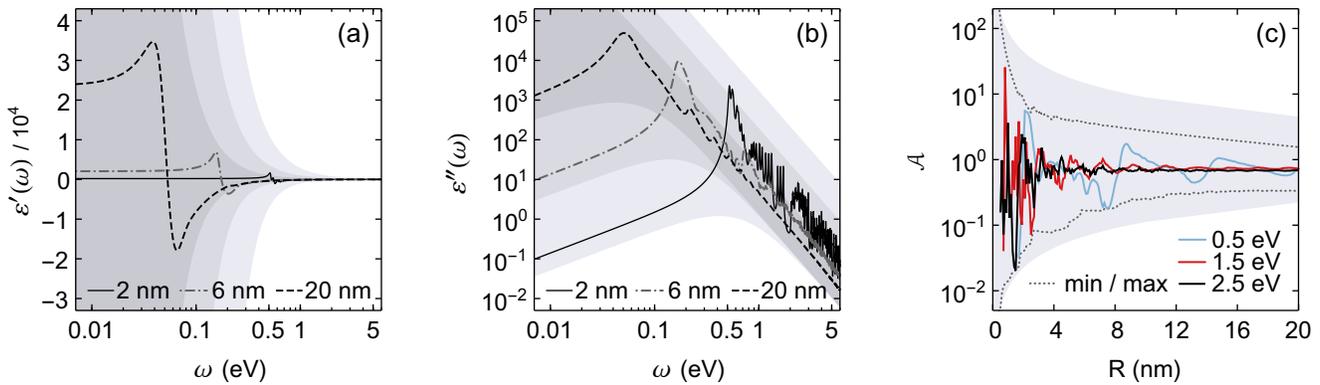}
\caption{\label{fig3} Bounds on (a) the real part of the QBM permittivity, (b) the imaginary part of the QBM permittivity, and (c) the size parameter extracted from the QBM permittivity using Eq. (\ref{eq:effective_size_param}). The shaded regions in (a) represent the bounds given by Eq.~(\ref{eq:bounds_real}), and the shaded regions in (b) represent the bounds in Eq.~(\ref{eq:bounds_imag}), ranging from $R = 2 $ nm (lightest) to $R = 20$ nm (darkest). The shaded region in (c) is given by Eq. (\ref{eq:bounds_size_param}). All quantities were evaluated for silver particles.
}
\end{figure*}
At this point, the limits in Eq. (\ref{eq:sum_limits}) can be used to evaluate Eq.~(\ref{eq:eps_mcmahon}) for finite systems. However, the McMahon approximation is only valid when $n_F \gg 1$ (see Appendix A), so we only consider the limiting case of large $n_F$, which is also satisfied by our restriction $R >$ 1 nm. When $n_F$ is large,  resonances with the same value of $\Delta n$ begin to cluster together. As $n_F$ increases, these individual resonances merge into collective resonances located at each group's average frequency $\omega_{\Delta n}$ found by summing over quantum numbers other than $\Delta n$
\begin{equation}
	\label{eq:group_frequency}
    \omega_{\Delta n} \equiv \frac{1}{S_{\Delta n}}  \sum_{\mathclap{n, l, \Delta l}}{S_{n, l, \Delta n, \Delta l} \omega_{n, l, \Delta n, \Delta l}} = \omega_0 (2\Delta n + 1)
\end{equation}
where $\omega_0 = \left(\pi/2\right) v_F/R$. The group strength is found in a similar manner
\begin{equation}
\label{eq:collective_sif}
    S_{\Delta n} \equiv \sum_{\mathclap{n,l,\Delta l}}{S_{n,l,\Delta n,\Delta l}} = \frac{8}{\pi^2(2\Delta n + 1)^2}\text{.}
\end{equation}
Note that the group oscillator strengths $S_{\Delta n}$ satisfy the sum 
rule $\sum_{\Delta n=0}^{\infty}{S_{\Delta n}}=1$, as they must. The specific details on the derivation of the group frequencies Eq. (\ref{eq:group_frequency}) and group oscillator strengths Eq. (\ref{eq:collective_sif}) are shown in Appendix A.  
For $n_F \gg 1$, the permittivity thus acquires the form of a single sum over Lorentzian resonances:
\begin{equation}
\label{eq:approx_sum}
    \varepsilon_{\infty}(\omega, R) = 1 + \omega_{p}^2 \sum_{\Delta n = 0}^\infty{ \frac{S_{\Delta n}}{ \omega_{\Delta n}^2 - \omega^2 - i\omega\gamma_0 } } \text{.}
\end{equation}
The permittivity given by Eq. (\ref{eq:approx_sum}) has a closed-form solution 
$\varepsilon_{\infty}(\omega, R) = \varepsilon_{D}(\omega) + \varepsilon_{s}(\omega, R)$, 
where $\varepsilon_D$ is the Drude permittivity [Eq.~(\ref{eq:drude_fn})], and $\varepsilon_{s}$ is a finite-size contribution given by
\begin{equation}
	\label{eq:closed_form_eps}
    \varepsilon_{s}(\omega, R) = \frac{2 \omega_{p}^2 \omega_0}{\pi \tilde{\omega}^3}
        \tan{\left( \frac{\pi \tilde{\omega}}{ 2 \omega_0 } \right)}
\end{equation}
where $\tilde{\omega} \equiv \sqrt{\omega(\omega + i \gamma_0)}$. This result contains a continuum of resonances that broaden the lineshape of the optical response in a complex way. This exemplifies how applying the free path correction [Eq.~(\ref{eq:sizeparam})] to the bulk Drude function does not capture all of the relevant physics for very small particles. For larger objects, the finite-size contribution to the permittivity diminishes as $\varepsilon_s \sim 1/R$, and the permittivity is dominated by Drude behavior.

Although the result $\varepsilon_{\infty}(\omega, R)$ is an approximate solution to Eq.~(\ref{eq:exactsum}) for large particles, it can serve as the basis for estimating bounds on the permittivity due to finite-size effects since it was constructed by clustering each $\Delta n$ transition group into a single Lorentzian. Thus we assert that the minima and maxima of $\varepsilon_{\infty}(\omega, R)$ serve as true bounds on the original, unclustered band profile. This can be easily checked by a parametric sweep.

Bounds for the real part of the permittivity can be found by minimizing and maximizing the real part of Eq.~(\ref{eq:closed_form_eps}). Written in terms of the Drude susceptibility $\chi_D'(\omega) = \varepsilon_D'(\omega) - 1$, we find
\begin{equation}
\label{eq:bounds_real}
\begin{split}
    \chi_D'(\omega) \left( 1 + \frac{2 \omega_0}{\pi \omega}\csch{\left(\frac{\pi\gamma_0}{2\omega_0}\right)} \right)
    \leq  
    \varepsilon'(\omega, R) - 1
    \leq \\
    \chi_D'(\omega) \left(1 - \sqrt{1 + \frac{\omega_0^2}{\gamma_0^2} } \left( 1 + \frac{\pi^2\gamma_0^2}{12\omega_0}\right) \right)
\end{split}
\end{equation}
A lower bound for the imaginary part of the permittivity can be obtained by considering the $\Delta n = 0$ term in Eq.~(\ref{eq:approx_sum}), and the upper bound for the imaginary part is found by maximizing the imaginary part of Eq.~(\ref{eq:closed_form_eps}), 
\begin{equation}
\label{eq:bounds_imag}
\begin{split}
    \frac{8 \omega \gamma_0 \omega_{p}^2 / \pi^2}{(\omega_0^2 + \omega^2)^2 + (\omega \gamma_0)^2}
    & \leq  
    \varepsilon''(\omega, R)
    \leq \\
    \varepsilon_D''(\omega) &\left(1+ \frac{2 \omega_0}{\pi \gamma_0}\coth{\left(\frac{\pi \gamma_0}{ 4\omega_0}\right)} \right) \text{.}
\end{split}
\end{equation}
The detailed derivation of the bounds in Eqs. (\ref{eq:bounds_real}) and (\ref{eq:bounds_imag}) are provided in Appendix B. Figure ~\ref{fig3} shows that the numerical calculation of $\varepsilon(\omega, R)$ contains resonances that fluctuate several orders of magnitude, but the resonances remain within the bounds given above. This has also been verified numerically with a complete parametric sweep from 0 to 10 eV for $1 \text{ nm} \leq R \leq 20 \text{ nm}$.

To better compare with free path effect calculations, we also consider bounding behavior for high frequencies ($\omega \gg \gamma_0$). In this case, we can write both upper and lower bounds with Drude-like behavior (see Appendix B, Section 4)
\begin{equation}
\label{eq:bounds_imag_high_freq}
    1 + \frac{2 \omega_0}{\pi \gamma_0} \tanh{\left(\frac{\pi \gamma_0}{ 4\omega_0}\right)}
    \leq  
    \frac{\varepsilon''(\omega, R)} {\varepsilon_D''(\omega)}
    \leq
    1 + \frac{2 \omega_0}{\pi \gamma_0} \coth{\left(\frac{\pi \gamma_0}{ 4\omega_0}\right)} \text{.}  \nonumber
\end{equation}
Because the imaginary part of the Drude permittivity can be written $\varepsilon_D'' \approx \omega_p^2 \gamma/\omega^3$ for $\omega \gg \gamma_0$, we can write the high-frequency bounds as $\omega_{p}^2 \bar{\gamma}_{\text{L}}/\omega^3
    \leq \varepsilon''(\omega, R)
    \leq
    \omega_{p}^2 \bar{\gamma}_{\text{U}}/\omega^3$
where we have introduced the upper and lower bounds of the effective
relaxation frequency
\begin{align}
	\label{eq:gamma_bound_lower}
    \bar{\gamma}_{\text{L}} &= \gamma_0 + \frac{v_F}{R}\tanh{\left( \frac{\gamma_0 R}{2 v_F} \right)} \\
    \label{eq:gamma_bound_upper}
    \bar{\gamma}_{\text{U}} &= \gamma_0 + \frac{v_F}{R}\coth{\left( \frac{\gamma_0 R}{2 v_F} \right)}
\end{align}
Comparing with Eq. (\ref{eq:sizeparam}), we can also write Eqs. (\ref{eq:gamma_bound_lower}) and (\ref{eq:gamma_bound_upper}) as bounds on the size parameter
\begin{align}
	\label{eq:bounds_size_param}
	\tanh{\left( \frac{\gamma_0 R}{ 2 v_F} \right)} \leq \mathcal{A}(R) \leq \coth{\left( \frac{\gamma_0 R}{ 2 v_F} \right)}.
\end{align}
These analytical bounds are compared with the exact QBM calculation in Fig.~\ref{fig3}(c) where effective values of $\mathcal{A}$ have been extracted from $\varepsilon''(\omega, R)$ using Eq. (\ref{eq:effective_size_param}) at three different frequencies. The values fall within the shaded area, which represent the high-frequency bounds given by Eq.~(\ref{eq:bounds_size_param}). The tightest possible bounds for $\mathcal{A}(R)$ are the minimum and maximum values of $\mathcal{A}(\omega, R)$ for the frequency range $[\omega_1(R), \omega_M]$, where the lower limit $\omega_1(R)$ indicates the first transition frequency for a given particle radius, and the upper limit $\omega_M = 10$ eV is chosen to be sufficiently large such that $\omega_M \gg \omega_1(R)$ for $R > 1$~nm. Clearly, the true value of $\mathcal{A}$ is frequency dependent and fluctuates between the minimum and maximum values indicated by the dotted lines in Fig.~\ref{fig3}(c). The range of values for $\mathcal{A}(R)$ becomes significantly wide even for $R \approx 5 \text{ nm}$ and continues to widen for smaller particle sizes, demonstrating that the semiclassical model is highly inaccurate for metal nanoparticles with diameters less than 10 nm. For particles in this size range, a quantum-mechanical result like Eq.~(\ref{eq:exactsum}) is more appropriate.

The fluctuations visible in Figs. \ref{fig:permittivity_size_range} and \ref{fig3} are similar to the oscillatory behavior reported by others.\cite{Yannouleas:1993vy,Molina:2002wf,Molina:2003ks,Weick:2005fz,Govyadinov:2011ev,Li:2013hf} In experiments with inhomogeneous broadening effects such as size dispersion or surface roughness, these fluctuations may be smoothed out sufficiently that it is suitable to use average values of the optical functions. But for measurements performed on highly uniform particle samples or on individual nano-objects, this highly oscillatory behavior will remain. In these situations, bounds like the ones shown in Fig. \ref{fig3} can be used to estimate a range of possible values.

\section{\label{sec:comparison_with_experiment}Comparison with experiment}

From the previous two sections, it is clear that quantum confinement can lead to large fluctuations in the permittivity and relaxation rates of finite-sized metal nanoparticles. Such large fluctuations frequently go undetected experimentally due to the presence of additional inhomogeneous broadening effects in existing experimental techniques. In this section, we study how a nonuniform size distribution of particles can suppress these fluctuations and how differences in sample dispersity may explain discrepancies between different experiments. 

In what follows, we use published experimental data for silver nanoparticles embedded in glass. \cite{Hovel:1993jo,Hilger:2001uz} The experiments consider the broadening of the surface plasmon resonance in the absorption spectra and extract the effective permittivity of the particles. Because the experimental samples are not uniform in size, we must perform averaging over a proper size distribution function $f(r,R,\sigma)$ with mean particle radius $R$ and standard deviation $\sigma$. The effective particle permittivity $\bar{\varepsilon}_p$ is then obtained through the averaged polarizability using the Maxwell-Garnett theory 
\begin{equation}
\label{eq:sample_permittivity}
	\frac{\bar{\varepsilon}_p(\omega, R, \sigma) - \varepsilon_m}{\bar{\varepsilon}_p(\omega, R, \sigma) + 2\varepsilon_m} = \int_0^\infty{f(r, R, \sigma) \frac{\varepsilon_p(\omega, R) - \varepsilon_m}{\varepsilon_p(\omega, R) + 2\varepsilon_m} dr}
\end{equation}
where $\varepsilon_{p}(\omega, R)$ is the permittivity of a particle with fixed radius $R$, and $\varepsilon_m$ is the permittivity of the embedding medium. Since the experimental data provides the values of the bound electron contribution to the permittivity $\varepsilon_b$, we can use Eq. (\ref{eq:sample_permittivity}) to extract the effective conduction electron permittivity $\varepsilon_D(\omega, R) = \varepsilon_p(\omega, R) - \varepsilon_b$ and hence the relaxation rate and size parameter according to Eq. (\ref{eq:effective_size_param}). The values of the bound electron permittivity according to each experiment are shown in Table \ref{tab:eps_b}.

\begin{table}[b]
\caption{\label{tab:eps_b}
Values used for interband corrections, taken from experimental measurements on bulk silver at the resonance frequency for silver in glass ($\omega_{sp} = 3.12$ eV)
}
\begin{ruledtabular}
\begin{tabular}{lccr}
\textrm{Ref.}&
\textrm{$\varepsilon_b'$}&
\textrm{$\varepsilon_b''$}\\
\colrule
Johnson and Christy \cite{Christy:1972di} & $4.19$ & 0.15 \\
Kreibig \textit{et al.} \cite{Hovel:1993jo} & $4.38$ & 0.24 \\
Hilger \cite{Hilger:2001uz} & $4.20$ & 0.90 \\
\end{tabular}
\end{ruledtabular}
\end{table}

\begin{figure}
\includegraphics{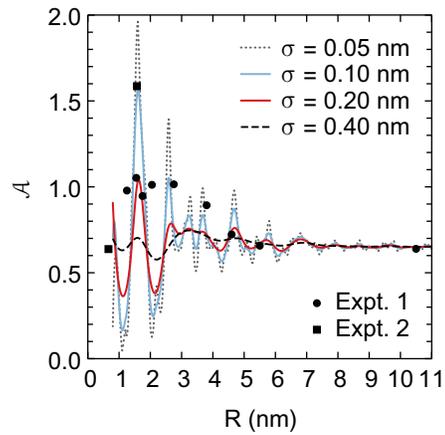}
\caption{
\label{fig:comparison_with_experiment}
The size parameter of silver nanoparticles embedded in glass ($\varepsilon_m = 2.3$). Theoretical values (lines) were calculated for normal size distributions using the QBM with effective medium theory [see Eq. (\ref{eq:sample_permittivity})]. Values obtained from experimental data is also shown from Ref. [\onlinecite{Hovel:1993jo}] (Expt. 1) and  Ref. [\onlinecite{Hilger:2001uz}] (Expt. 2). All values were calculated at the resonance frequency of silver particles in glass ($\omega_{sp} = 3.12$ eV).
}
\end{figure}

Fig. \ref{fig:comparison_with_experiment} shows the experimental data compared with calculations of normally distributed particle samples using $f(r, R, \sigma) = \exp\textbf{(}-(r - R)^2/(2\sigma^2)\textbf{)} / \sqrt{2\pi\sigma^2}$. The values were calculated at the surface plasmon frequency of silver nanoparticles in glass in correspondence with the experimental conditions ($\omega_{sp} = 3.12$ eV). The calculations predict significant fluctuations in the size parameter for small particles ($R < 5$ nm), whereas the size parameter quickly collapses to a constant value for larger particle sizes. The size-dependent fluctuations disappear almost entirely even for a relatively narrow size distribution with $\sigma = 0.4$~nm. The experimental values fall within the range of values predicted by the QBM. Varying the dispersity ($\sigma$) of the sample can also explain the variation in the data points.

The usage of the finite-size correction in Eq.~(2) is widespread. Figure \ref{fig:comparison_with_experiment} provides insight into its applicability for many realistic samples of metal particles since most current experimental techniques are limited to measurements on a sample with varying size, shape, and orientation. The necessity of quantum corrections to Eq.~(2) depends strongly on both the average size and the dispersity of the sample.  Because nonuniform particle samples suppress almost all quantum effects, extremely narrow size distributions are required to reliably test the predictions of the QBM.

\section{\label{sec:conclusion}Conclusion}
In this paper, we performed a detailed study of size effects in the permittivity of metal nanoparticles using the QBM. By deriving strict analytical bounds on the permittivity, relaxation rates, and semiclassical size parameter $\mathcal{A}$, we investigated the limits to which quantum confinement effects can enhance their optical properties. We also argue that caution should be exercised when following the common practice of modeling finite-sized systems with a modified Drude function and Eq.~(\ref{eq:sizeparam}). Quantum effects should be accounted for in uniform samples containing particles with $R \lesssim 10$ nm by either (i) using a quantum-mechanical permittivity like Eq. (\ref{eq:exactsum}), (ii) studying proper bounds on the permittivity like Eqs. (\ref{eq:bounds_real}) and (\ref{eq:bounds_imag}), or (iii) introducing a size-dependent and frequency-dependent function $\mathcal{A}(\omega, R)$.

Finally, by comparing the theory with experimental data, we provide an example of how a non-uniform size distribution can suppress the effects of quantum confinement. This can also explain the variance in measured values of the Drude size parameter $\mathcal{A}$, even when the measurements are performed under similar experimental conditions. Future experiments on highly uniform particle samples or even single nano-objects are needed to adequately test the size-dependent oscillatory behavior predicted by the QBM.

\appendix\label{appendix:sum_rule}

\section{Evaluation of the McMahon sum rule}
The exact sum rule $\sum_{ss'}{S_{ss'}} = 1$ is quite generally valid; however, the asymptotic McMahon approximation $a_{nl} \approx \pi(n + 1 + l/2)$ is only accurate when $n \gg l$, and it is not obvious that the sum rule is satisfied under this approximation. Here we demonstrate that the sum rule is indeed satisfied in the limit of large particle sizes ($n_F \gg 1$).

We start by noting that the limits in Eqs. (\ref{eq:sum_limits}a)--(\ref{eq:sum_limits}d) are the same for the two $\Delta l = \pm 1$ cases if we apply the transformation $\Delta n \rightarrow \Delta n + 1$ when $\Delta l = -1$ and assume $n_F \gg 1$. We can thus combine both $\Delta l$ terms and use Eq. (\ref{eq:approx_sif}) to define a combined strength factor $T_{n,l,\Delta n} = S_{n,l,\Delta n + 1,\Delta l = -1} + S_{n,l,\Delta n,\Delta l = +1}$ given by
\begin{align}
	T_{n, l, \Delta n} &\equiv \frac{16 (2 n_F - l + 2)^2}{\pi^2 n_F^3 (2\Delta n + 1)^3 } \nonumber \\
	&\times \frac{ (4 n_F - 2 l- 4 n + 1) (2 \Delta n-l+2 n_F+3)^2}{(2 \Delta n - 2 l+4 n_F+5)^3} \nonumber \text{.}
\end{align}
In writing $T_{n, l, \Delta n}$, we also applied the simplifying transformation $l \rightarrow -l + 2(n_F-n)$ and used the relation $N_s = (4/3)n_F^3$, which is the correct density of states for the McMahon approximation. Others have pointed out \cite{Hache:1986tx,Barma:1989tx,Broglia:1992vi} that the McMahon density of states does not agree with the bulk density of states, but we use the McMahon relation so that the model remains self-consistent.

\begin{figure}
\includegraphics{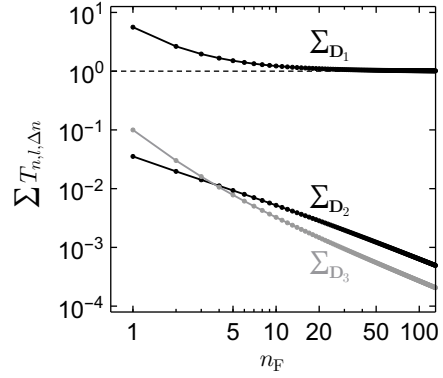}
\caption{
\label{fig:appendix_sum_rule}
Evaluation of the sum rule for the three different domains defined in Appendix A.
}
\end{figure}

We define the set $\mathbf{D}(n, l, \Delta n) = \{n, l, \Delta n\}$ as the set of values $n$, $l$, and $\Delta n$ which satisfy the summation constraints in Eqs. (\ref{eq:sum_limits}a)--(\ref{eq:sum_limits}d) with the transformations described in the previous paragraph. With these conditions, we can write the set $\mathbf{D}$ as the union of three subsets, $\mathbf{D} = \mathbf{D}_1 \cup \mathbf{D}_2 \cup \mathbf{D}_3$, where we define the subsets as follows:
\small
\begin{align*}
	\hline
	\hline
	&   \scalebox{1.1}{$\mathbf{\mathbf{D}}_1$} && \scalebox{1.1}{$\mathbf{D}_2$} && \scalebox{1.1}{$\mathbf{D}_3$}  \rule{0pt}{2.6ex} \rule[-0.9ex]{0pt}{0pt} \\
	\hline
	&  0 \leq \Delta n \leq n_F     &\hspace{0.2em} &   0 \leq \Delta n \leq n_F        &\hspace{0.2em} &  \Delta n > n_F            \rule{0pt}{2.6ex} \\
	&  0 \leq n \leq n_F - \Delta n &\hspace{0.2em} &   n_F - \Delta n < n \leq n_F  &\hspace{0.2em} &  0 \leq n \leq n_F         \\
	&  0 \leq l \leq 2 \Delta n     &\hspace{0.2em} &   0 \leq l \leq 2 (n_F - n)       &\hspace{0.2em} &  0 \leq l \leq 2 (n_F - n) \\[0.5em]	
	\hline
	\hline
\end{align*}
\normalsize
The sum rule can then be written $\sum_{ss'}{S_{ss'}} = \sum_{\mathbf{D}}{T_{n,l,\Delta n}}=\sum_{\mathbf{D_1}}{T_{n,l,\Delta n}} + \sum_{\mathbf{D_2}}{T_{n,l,\Delta n}} + \sum_{\mathbf{D_3}}{T_{n,l,\Delta n}}$. From Fig.~\ref{fig:appendix_sum_rule}, it's clear that the sums over $\mathbf{D_2}$ and $\mathbf{D_3}$ vanish when $n_F$ is large, so we have $\sum_{\mathbf{D}}{T_{n,l,\Delta n}} \simeq \sum_{\mathbf{D}_1}{T_{n,l,\Delta n}}$ for large $n_F$. If we write the upper summation limit for $n$ assuming that $n_F \gg \Delta n$, then we find the group oscillator strengths
\begin{align}
	S_{\Delta n} \equiv \lim_{n_F \to \infty} \sum_{l=0}^{\mathclap{2\Delta n}} \sum_{n=0}^{n_F}{T_{n, l, \Delta n}}  &= \sum_{\mathclap{l=0}}^{2\Delta n} \frac{8}{\pi^2 (2\Delta n + 1)^3} \nonumber \\
		&= \frac{8}{\pi^2 (2\Delta n + 1)^2}
\end{align}
where we first performed the straightforward sum over $n$ and subsequently applied the limit $n_F \to \infty$. The final sum over $l$ can then be evaluated easily. The sum rule $\sum_{\Delta n = 0}^{\infty}{S_{\Delta n}} = 1$ is readily verified, confirming that the sum rule is satisfied for $n_F~ \gg 1$.

The group frequencies are evaluated analogously to the sum rule. Combining $\omega_{n,l,\Delta n,\Delta l}$ for both $\Delta l = \pm 1$ terms and again applying the transformation $l \rightarrow -l + 2(n_F-n)$, we find
\begin{align*}
	\Omega_{n, l, \Delta n} &\equiv \frac{\hbar \pi^2}{8 M R^2} (2 \Delta n + 1)   (2 \Delta n - 2 l+4 n_F + 5)
\end{align*}
For large $n_F$ we again only need to consider the sum corresponding to $\mathbf{D_1}$, and the group average frequency becomes
\begin{align*}
	\omega_{\Delta n} \equiv \frac{1}{S_{\Delta n}}  \sum_{l=0}^{\mathclap{2\Delta n}} \sum_{n=0}^{n_F}{T_{n, l, \Delta n}\Omega_{n, l, \Delta n}}
\end{align*}
Keeping only the leading $n_F$ term, we find
\begin{align}
	\omega_{\Delta n} = \frac{\hbar \pi^2 n_F}{2 M R^2}(2\Delta n + 1) = \omega_0 (2\Delta n + 1)
\end{align}
where we have defined the characteristic frequency $\omega_0 \equiv \hbar \pi^2 n_F / 2 M R^2 = (\pi/2) v_F / R$.

\section{Bounds on the permittivity}
We are ultimately interested in bounds for $\varepsilon_{\infty}(\omega, R)$. We first recast the problem in a more illuminating form by extracting the Drude term from the sum in Eq. (\ref{eq:approx_sum}). We can then split the finite-size contribution into its real and imaginary parts by writing
\begin{align}
\label{eq:split_fn}
    \text{Re}\left[\varepsilon_{\infty}(\omega)\right] - 1 &= \chi_D' (1 + \xi(\nu, \tau)) \\
    \text{Im}\left[\varepsilon_{\infty}(\omega)\right] &= \chi_D'' (1 + \eta(\nu, \tau))
\end{align}
where we have defined the parameters $\nu = \omega/\omega_0$ and $\tau = \gamma/\omega_0$. In this form, the scaling functions $\xi(\nu, \tau)$ and $\eta(\nu, \tau)$ represent corrections to the Drude susceptibility functions $\chi_D'$ and $\chi_D''$. The scaling functions are given by
\begin{align}
\label{eq:split_fn_xi}
    \xi(\nu, \tau) &= \frac{8}{\pi^2} \sum_{n=0}^{\infty}{ \frac{\nu^2 - \tau^2 - 1 - 4n(n+1)}{((2n+1)^2 - \nu^2)^2 + \nu^2\tau^2} } \\
\label{eq:split_fn_eta}
    \eta(\nu, \tau) &= \frac{8}{\pi^2} \sum_{n=0}^{\infty}{ \frac{2\nu^2-(2n+1)^2}{((2n+1)^2 - \nu^2)^2 + \nu^2\tau^2} }
\end{align}
The problem is now reduced to finding bounds for the functions $\xi$ and $\eta$. The summations in Eqs. (\ref{eq:split_fn_xi}) and (\ref{eq:split_fn_eta}) have the closed-form solutions $\xi(\nu, \tau) = -\text{Re}[f(\nu, \tau)] / \nu$ and $\eta(\nu, \tau) = \text{Im}[f(\nu, \tau)] / \tau$ where

\begin{align}
	\label{eq:closed_form_f}
    f(\nu, \tau) &= \frac{2 \thinspace (\nu - i \tau)^{3/2} \tan{ \left( \frac{\pi}{2} \sqrt{\nu(\nu + i \tau)} \right) } }{\pi \sqrt{\nu (\nu^2 + \tau^2)}}
\end{align}
is a complex-valued function. To investigate the bounding behavior of these functions, we write Eq. (\ref{eq:closed_form_f}) entirely in terms of real-valued functions. Making use of the property $\tan{(x + i y)} = \left(\sin{(2x)} + i \sinh{(2y)} \right) / \left( \cos{(2x)} + \cosh{(2y)} \right)$, we can write $f(\nu, \tau) = -F(\nu, \tau) + i G(\nu, \tau)$ where

\begin{figure*}
\includegraphics{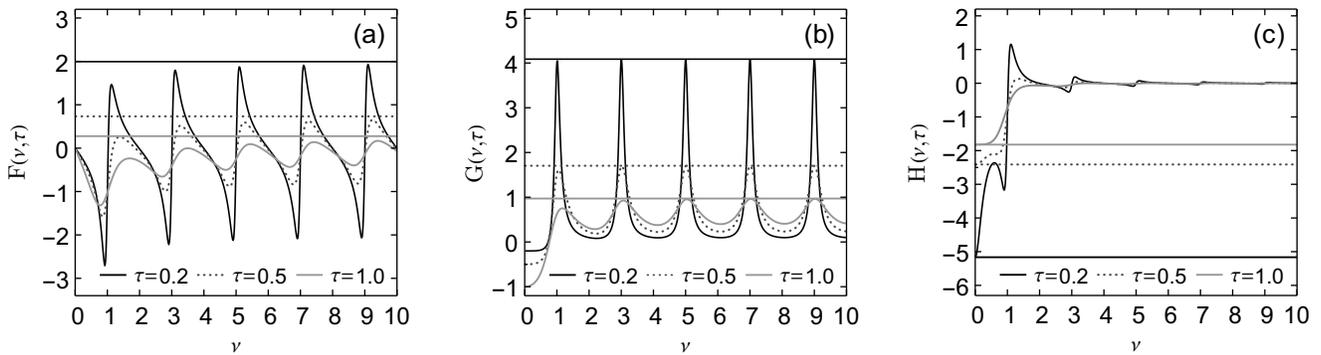}
\caption{
\label{fig:appendix_bound_functions}
The regularized scaling functions (a) $H(\nu, \tau) = (\nu\sqrt{\nu^2+\tau^2})^{-1}F(\nu, \tau)$, (b) $F(\nu, \tau)$, and (c) $G(\nu, \tau)$ as defined in Appendix B. Horizontal lines indicate limiting values. Global minima for $H(\nu, \tau)$ occurring at $\nu = 0$ are shown, and the global maxima of the functions $F(\nu, \tau)$ and $G(\nu, \tau)$ as $\nu \to \infty$ are shown for different values of $\tau$.
}
\end{figure*}

\begin{widetext}
\begin{align}
	\label{eq:split_fn_F}
    F(\nu, \tau) = \frac{ 2 Y (Y^2 - 3 X^2) \sinh(\pi \nu Y) - 2 X (X^2 - 3Y^2) \sin{(\pi \nu X)}} {\pi(X^2 + Y^2) (\cos{(\pi \nu X)} + \cosh{(\pi \nu Y))} } \\
    \label{eq:split_fn_G}
    G(\nu, \tau) = \frac{ 2 Y (Y^2 - 3X^2) \sin{(\pi \nu X)} + 2 X (X^2  - 3 Y^2) \sinh(\pi \nu Y)} {\pi(X^2 + Y^2) (\cos{(\pi \nu X)} + \cosh{(\pi \nu Y))} }
\end{align}
with $X \equiv (1/\sqrt{2}) \sqrt{1 + \sqrt{1 + (\tau/\nu)^2}}$ and $Y \equiv (1/\sqrt{2}) \sqrt{-1 + \sqrt{1 + (\tau/\nu)^2}}$.
\end{widetext}

The scaling functions are thus $\xi(\nu, \tau) = F(\nu, \tau) / \nu$ and $\eta(\nu, \tau) = G(\nu, \tau) / \tau$. We now use the results Eqs. (\ref{eq:split_fn_F}) and (\ref{eq:split_fn_G}) to seek upper and lower bounds on the scaling functions, i.e. $\xi^{-} \leq \xi \leq \xi^{+}$ and $\eta^{-} \leq \eta \leq \eta^{+}$. The corresponding bounds on the permittivity are as follows:
\begin{align}
	\label{eq:gen_bounds_real}
    \chi_D'\left(1 + \xi^{+} \right) &\leq \varepsilon' - 1 \leq \chi_D' \left(1 + \xi^{-} \right) \\
    \label{eq:gen_bounds_imag}
    \varepsilon''_D\left(1 + \eta^{-} \right) &\leq \varepsilon'' \leq \varepsilon_D'' \left(1 + \eta^+ \right)
\end{align}
Where possible, we explore frequency-independent bounds $\xi^{-}(\tau)$, $\xi^{+}(\tau)$, $\eta^{-}(\tau)$, and $\eta^{+}(\tau)$ to preserve the frequency dependence of the Drude functions in the bounds Eqs. (\ref{eq:gen_bounds_real}) and (\ref{eq:gen_bounds_imag}) above.
\subsection{Lower bound for $\xi(\nu, \tau)$}

It can be readily verified that the global minimum of $\xi(\nu, \tau)$ is always located in the range $0 \leq \nu_{\text{min}} \leq 1$. However, the exact value of $\nu_{\text{min}}$ depends on $\tau$ in a nontrivial way. We instead consider the function $H(\nu, \tau) \equiv (\nu^2 + \tau^2)^{-1/2} \xi(\nu, \tau) = (\nu\sqrt{\nu^2+\tau^2})^{-1}F(\nu, \tau)$ shown in Fig. \ref{fig:appendix_bound_functions}(a). This function is minimized at $\nu = 0$ for all values of $\tau$, and the minimum value can be found by taking the limit as $\nu$ goes to zero:
\begin{align}
\lim_{\nu \rightarrow 0}{(\nu^2 + \tau^2)^{-1/2} \xi(\nu, \tau)} = -\left(\frac{1}{\tau} + \frac{\pi^2 \tau}{12}\right) \nonumber
\end{align}
Since this is a minimum value, we can use it to write a frequency-dependent lower bound $\xi(\nu, \tau) \geq \xi^{-}(\nu, \tau)$ with
\begin{align}
	\label{eq:xi_bound_1}
	\xi^{-}(\nu, \tau) \equiv -(\nu^2 + \tau^2)^{1/2} \left( \frac{1}{\tau} + \frac{\pi^2 \tau}{12} \right) \text{.}
\end{align}
The bound $\xi^{-}(\nu, \tau)$ holds for all values of $\nu$, so it must also be true that $\xi(\nu, \tau) \geq \xi^{-}(\nu_{\text{min}}, \tau)$. Since we always have $\nu_{\text{min}} \leq 1$ for $\xi(\nu, \tau)$, we can also establish a frequency-independent bound by evaluating Eq. (\ref{eq:xi_bound_1}) at $\nu = 1$:
\begin{align}
	\xi^{-}(\tau) \equiv -(1 + \tau^2)^{1/2} \left( \frac{1}{\tau} + \frac{\pi^2 \tau}{12} \right) \text{.}
\end{align}

\subsection{Upper bound for $\xi(\nu, \tau)$}
Observe in Fig.~\ref{fig:appendix_bound_functions}(b) that the function $\nu \thinspace \xi(\nu, \tau) = F(\nu, \tau)$ has an infinite set of local maxima and minima which always monotonically increase to an asymptotic value. The limiting value can be found from the asymptotic behavior for $\nu \gg \tau$, in which case $X \approx 1$ and $Y \approx \tau / 2 \nu$. Dropping terms containing $\tau / \nu$, we obtain
\begin{align}
	\label{eq:xi_asymptotic}
    F(\nu, \tau) \sim \frac{-2 \sin{ \left( \pi \nu \right)} }{\pi \cos{ \left( \pi \nu \right)} + \pi \cosh{ \left( \pi \tau / 2 \right)}}, \quad (\nu \gg \tau).
\end{align}
This function has minima and maxima at $\nu_{\text{min,max}} = 2n \pm \arccos{ \textbf{(} -\sech{(\pi \tau / 2)} \textbf{)} }$. Evaluating Eq. (\ref{eq:xi_asymptotic}) at these points, we find that the function asymptotically oscillates between two limiting values,
\begin{align}
	\nonumber
    \lim_{\nu \rightarrow \infty}{F(\nu, \tau)} = \left[-\frac{2}{\pi} \csch \left(\frac{\pi \tau}{2} \right), \frac{2}{\pi} \csch \left(\frac{\pi \tau}{2} \right) \right]
\end{align}
Because the maxima of $F(\nu, \tau)$ increase monotonically, the positive asymptotic value must be an upper bound for all frequencies. Therefore, we have $\xi(\nu, \tau) \leq \xi^+ \equiv (2/\pi \nu) \csch \left(\pi \tau / 2 \right) $.

\subsection{Lower bound for $\eta(\nu, \tau)$}
The function $\eta(\nu, \tau)$ has a global minimum $\eta(0, \tau) = -1$. This establishes the bound $\varepsilon'' \geq \chi_D''$, which simply confirms that the imaginary part of the finite-size contribution $\varepsilon_s''$ is always positive. A tighter lower bound for $\varepsilon''$ can be found by instead considering the summation in Eq.~(\ref{eq:approx_sum}). Since each imaginary term is always positive, every $\Delta n$ term is a lower bound. We give the result for the first term ($\Delta n = 0$) in Eq.~(\ref{eq:bounds_imag}).

\subsection{Upper and high-frequency bounds for $\eta(\nu, \tau)$}
The function $G(\nu, \tau)$ has monotonically increasing maxima, so we again find the high-frequency asymptotic for $\nu \gg \tau$. Taking $X \approx 1$ and $Y \approx \tau / 2 \nu$, we find
\begin{align}
	\label{eq:eta_asymptotic}
    G(\nu, \tau) \sim \frac{2\sinh{ \left( \pi \tau / 2 \right)} }{\pi \tau \cos{ \left( \pi \nu \right)} + \pi \tau \cosh{ \left( \pi \tau / 2 \right)}}
\end{align}
Maximizing/minimizing with respect to $\nu$, we find that minima occur at $\nu_{\text{min}} = 2n$ and maxima at $\nu_{\text{max}} = 2n + 1$. The corresponding values are
\begin{align}
    \lim_{\nu \rightarrow \infty}{G(\nu, \tau)} = \left[\frac{2}{\pi} \tanh \left(\frac{\pi \tau}{4} \right), \frac{2}{\pi} \coth \left(\frac{\pi \tau}{4} \right) \right]
\end{align}
The lower value is a high frequency bound, but the upper value holds as a bound for all frequencies. Thus, we can conclude that $\eta(\nu, \tau) \geq \eta^-$ and $\eta(\nu, \tau) \leq \eta^+$ for all $\nu$ when $\nu \gg \tau$, where $\eta^- \equiv (2/\pi \tau) \tanh \left(\pi \tau / 4 \right)$ and $\eta^+ \equiv (2/\pi \tau) \coth \left(\pi \tau / 4 \right)$.

\begin{acknowledgments}
This paper was supported by the NSF EPSCoR CIMM project under Award No. OIA-1541079 and the Louisiana Board of Regents.
\end{acknowledgments}

\bibliography{ref}

\end{document}